\documentclass[10pt,twoside,letter,amsmath,amssymb,aps,showpacs]{revtex4}
\usepackage{amsfonts}
\usepackage{graphics}
\usepackage{epsfig}
\usepackage{color}

\begin{document}
\large

\thispagestyle{myheadings}

\title{High precision calculations of electroweak radiative corrections
for polarized M{\o}ller scattering at one loop and beyond}
\author{Aleksandrs Aleksejevs}
\email{aaleksejeve@grenfell.mun.ca}
\affiliation{Grenfell Campus of Memorial University, Corner Brook, Canada}

\author{Svetlana Barkanova}
\email{svetlana.barkanova@acadiau.ca}
\affiliation{Acadia University, Wolfville, Canada}

\author{Yury Kolomensky}
\email{yury@physics.berkeley.edu}
\affiliation{University of California, Berkeley, USA}

\author{Eduard Kuraev}
\email{kuraev@theor.jinr.ru}
\affiliation{Joint Institute for Nuclear Research, Dubna, Russia}

\author{Vladimir Zykunov}
\email{vladimir.zykunov@cern.ch}
\affiliation{Belarussian State University of Transport, Gomel, Belarus}



\begin{abstract}
Parity-violating M{\o}ller scattering measurements are a powerful probe of new physics effects, 
and the upcoming high-precision experiments will require a new level of accuracy for electroweak radiative corrections (EWC). 
First, we perform the updated calculations of one-loop EWC for M{\o}ller scattering 
asymmetry using two different approaches: semi-automatic, precise, with FeynArts and FormCalc as base languages, and  ``by hand'', 
with reasonable approximations. In addition, we provide a tuned comparison between the one-loop results obtained in two different 
renormalization schemes: on-shell and constrained differential renormalization. As the last step, we discuss the two-loop EWC induced by 
squaring one-loop diagrams, and show that the significant size of this partial correction indicates 
a need for a complete study of the two-loop EWC in order to meet the precision goals of future experiments.
\end{abstract}


\maketitle

\section{Introduction}
M{\o}ller scattering is a very clean process with well-known kinematics and extremely suppressed backgrounds,
 and any inconsistency with the Standard Model will signal new physics. The next-generation experiment to study
 electron-electron scattering, MOLLER \cite{MOLLER}, planned at JLab following the 11 GeV upgrade, will offer a new 
level of sensitivity and measure the parity-violating asymmetry in the scattering of longitudinally polarized electrons
 off an unpolarized target to a precision of 0.73 ppb, and allow a determination of the weak mixing angle with an uncertainty 
of about 0.1\%, a factor of five improvement over the measurement by E-158 \cite{E158}. Obviously, 
before we can extract reliable information from the experimental data, it is necessary to take into account EWC. EWC to the parity-violating 
(PV) M{\o}ller scattering asymmetry were addressed in the literature 
earlier and were shown to be large~\cite{CzarMarc,DenPozz,ABIZ}. A more detailed literature review can be found in \cite{ ABIZ},
 our first work on the topic. In \cite{ABIZ}, we calculated a full gauge-invariant set of the one-loop EWC and found the total
 correction to be close to $-70$\%, with no significant theoretical uncertainty coming from the hadronic contributions to the vacuum
 polarization or other uncertain input parameters. Since it is possible that a much larger theoretical uncertainty may come from 
two-loop corrections, we investigated the importance of two-loop contribution in \cite{ABIKZ},
 by comparing corrections calculated in two different renormalization schemes (RS) -- 
on-shell (OS) and constrained differential  renormalization (CDR) \cite{CDR} -- 
and found a difference of about 11\%. This means that the two-loop EWC may be larger than previously
 thought and cannot be dismissed, especially in the light of precision promised by MOLLER. We divide the two-loop EWC into two
 classes: the Q-part induced by quadratic one-loop amplitudes, and the T-part which includes the interference of Born and two-loop diagrams.
 In \cite{q-part}, we calculated the Q-part exactly and found that it can reach 4\%. Here, we provide a
 brief review of our calculations done at the one-loop level \cite{ ABIZ}, show details of the comparison between the corrections
 evaluated in the OS and CDR schemes \cite{ABIKZ}, and outline some of our calculations of higher order corrections.

\section{Born and one-loop corrections}
\begin{figure}[h]
\begin{centering}
\includegraphics[scale=0.32]{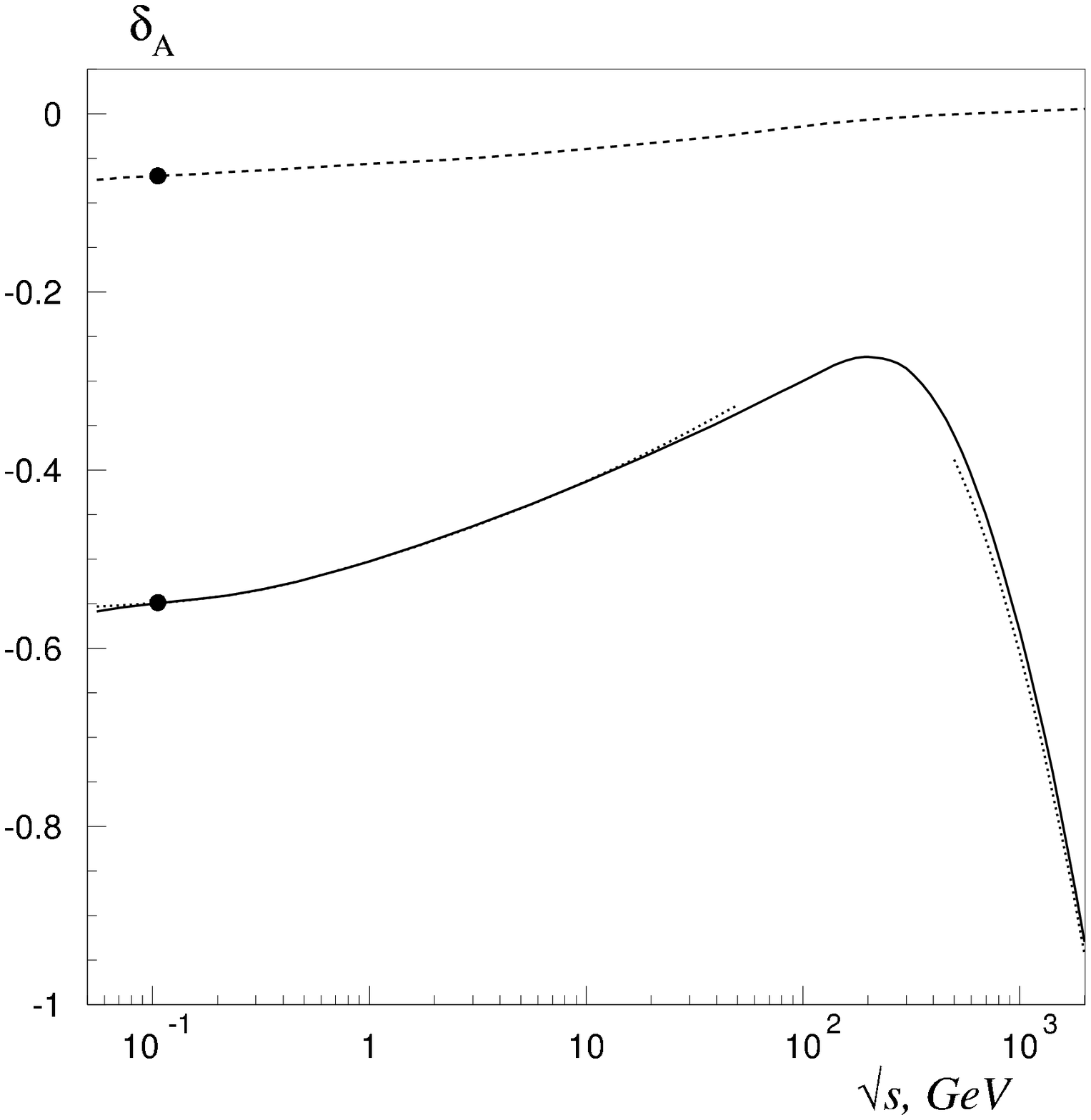}~
\includegraphics[scale=0.32]{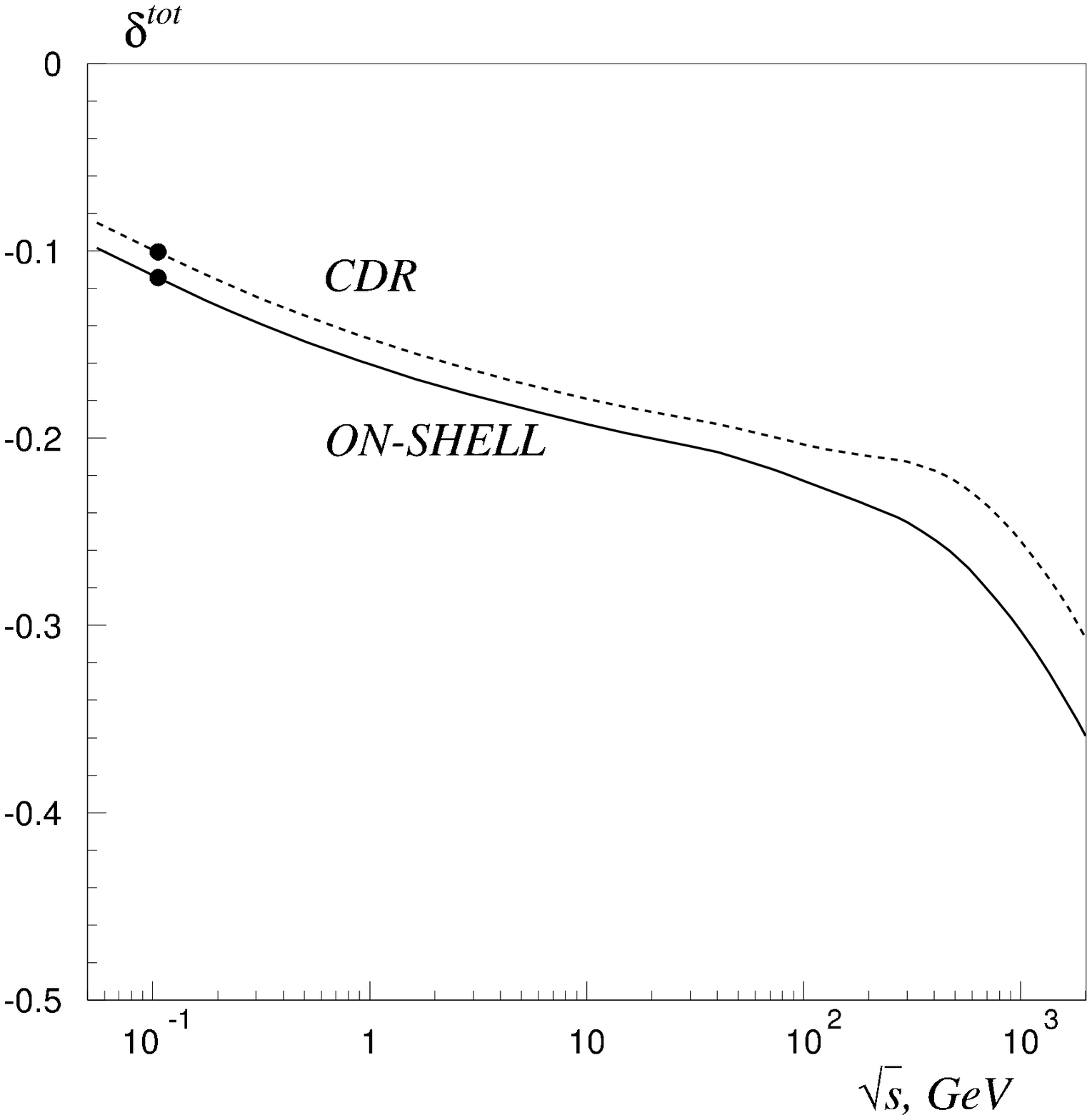}~
\includegraphics[scale=0.325]{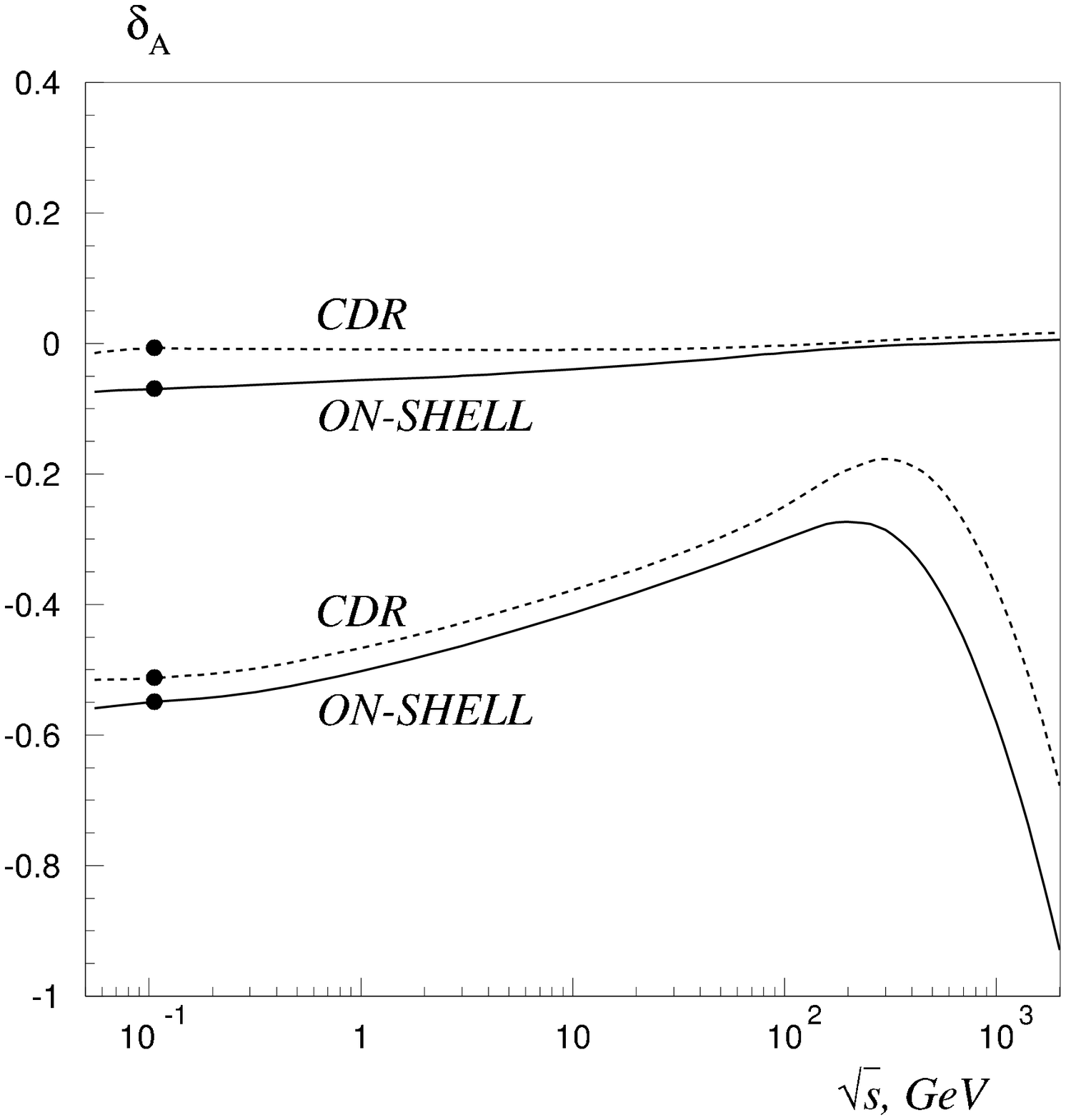}
\par\end{centering}

\caption{a) Weak (solid) and QED (dashed) ($\omega=0.05\cdot\sqrt{s}$) corrections to the PV
asymmetry in M{\o}ller scattering at $\theta=90^{\circ}$ (left plot).
b) Total correction to the cross section in the OS and CDR schemes (middle
plot). c) Correction to the PV asymmetry in  the OS and CDR schemes (right
plot)}
\label{one-loop-cor-asym}

\end{figure}

The asymmetry between left/right longitudinally polarized electrons can be constructed in the following
way:
\begin{eqnarray}
A_{LR}=\frac{\sigma_{LL}+\sigma_{LR}-\sigma_{RL}-\sigma_{RR}}{\sigma_{LL}+\sigma_{LR}+\sigma_{RL}+\sigma_{RR}}=\frac{\sigma_{LL}-\sigma_{RR}}{\sigma_{LL}+2\sigma_{LR}+\sigma_{RR}},\label{eq:a1}
\end{eqnarray}
enhancing the contributions induced by PV electroweak interactions. The term $\sigma\equiv\frac{d\sigma}{d\cos\theta}$ stands for
the differential cross section defined in the center of mass reference
frame of incoming electrons. At the Born level (leading order
(LO)), the asymmetry is
\begin{eqnarray}
A_{LR}=\frac{s}{2m_{W}^{2}}\frac{y(1-y)}{1+y^{4}+(1-y)^{4}}\frac{1-4s_{W}^{2}}{s_{W}^{2}}, & y=-\frac{t}{s},\label{eq:a2}
\end{eqnarray}
where $s_{W}^{2}\equiv\sin^{2}\theta_{W}=1-\frac{m_{W}^{2}}{m_{Z}^{2}}\sim0.24$. As one can see from Eq.~\ref{eq:a2}, the asymmetry is highly sensitive
to $\theta_{W}$ so any
deviation from the SM value will signal new physics. Obviously, before we can extract reliable information from the experimental data, it is necessary to include EWC. The cross section including one-loop matrix elements is:
\begin{eqnarray}
\sigma=\frac{\pi^{3}}{2s}|M_{0}+M_{1}|^{2}=\frac{\pi^{3}}{2s}\Bigl(M_{0}M_{0}^{\dagger}+2\mbox{Re}M_{1}M_{0}^{\dagger}+M_{1}M_{1}^{\dagger}\Bigr)=\sigma_{0}+\sigma_{1}+\sigma_{Q}\label{eq:a3}
\end{eqnarray}
where $\sigma_{1}=\sigma_{1}^{BSE}+\sigma_{1}^{Ver}+\sigma_{1}^{Box}\propto\alpha^{3}$
is an interference term between the Born and one-loop amplitudes (NLO), and the  cross section
$\sigma_{Q}\propto\alpha^{4}$ is a quadratic term of the
same order as the two-loop contribution (NNLO). To make sure that our calculations at the one-loop level are error-free, we evaluate EWC using
two different methods.
Our first method,   ``by hand'', is to derive the compact
analytic expressions for the leading one-loop correction manually
 using appropriate approximations 
for $\sqrt{s} < 30$~GeV and $\sqrt{s} > 500$~GeV~\cite{ABIZ}. 
Our second method, semi-automated, is to consider a full set of graphs with no
approximations using computer-based algebra packages \cite{FA,FC}
and \cite{form}. To make sure that we calculate a gauge-invariant
set of graphs, we use two sets of renormalization conditions (RC): the RC by Hollik
(HRC) introduced in \cite{HRC} for our  ``by hand'' approach, and the RC proposed by Denner (DRC) in \cite{DRC} for our semi-automated method. 
The infrared divergences (IR) are treated
by the soft and hard-photon bremsstrahlung (see \cite{ABIZ}). 
We choose our input parameters to be the fine structure constant ($\alpha=1/137.03599$), the
mass of the W boson ($m_{W}=80.398\,\mbox{GeV}$) and the mass of the
Z boson ($m_{Z}=91.1876\,\mbox{GeV}$). A relative correction to the PV asymmetry is defined as
$\delta_{A}^{C}=(A_{LR}^{C}-A_{LR}^{0})/{A_{LR}^{0}}$, 
with the superscript in $\delta_{A}^{C}$ corresponding
to the various contributions: ``weak'' indicates no IR-divergent graphs,
and ``QED'' indicates only IR divergent graphs treated by bremsstrahlung
contribution. In order to see how our results compare to the literature
\cite{DenPozz}, we compare the $\delta_{A}^{weak}(\mbox{[5]})=-0.2790$
for the $\sqrt{s}=100\,\mbox{GeV}$ and $\delta_{A}^{weak}(\mbox{[4]})=-0.2787$
using the same input parameters as in \cite{DenPozz} and obtain excellent agreement.
A comparison of results evaluated by two
methods can be seen on Fig. \ref{one-loop-cor-asym}(a).
%
To establish if the higher-order (NNLO) contributions in a
given RS are important (see \cite{HolTim}), we compare results
in two RS: OS and CDR. Fig.~\ref{one-loop-cor-asym}(b) shows
the total correction to the unpolarized cross section $\delta^{tot}=(\sigma^{tot}-\sigma^{0})/\sigma^{0}$
calculated in the OS and CDR schemes. In the low-energy region, the correction to the cross section is dominated by the
QED contribution, and the difference between the two schemes is almost constant and rather
small ($\sim0.01$), but it grows at $\sqrt{s}\geq m_{Z}$ as the weak correction
becomes comparable to QED. As a result (see Fig. \ref{one-loop-cor-asym}(c)), the difference between the
OS and CDR corrections to the PV asymmetry can reach as much as $10\%$, so contributions from two-loop corrections could
become important. 

\section{Two-loops corrections: Q-part}
\begin{figure}
\begin{centering}
\includegraphics[scale=0.5]{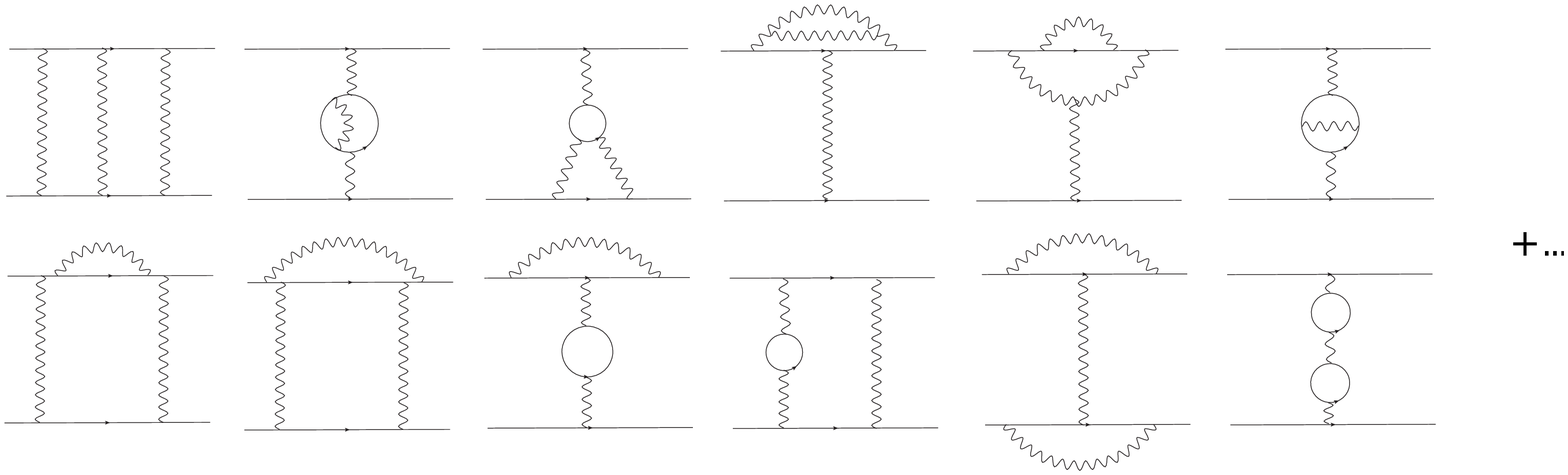}
\par\end{centering}

\caption{Representative two-loop graphs for the M{\o}ller scattering.}
\label{two-loops-Moller}
\end{figure}

\begin{figure}
\begin{centering}
\includegraphics[scale=0.5]{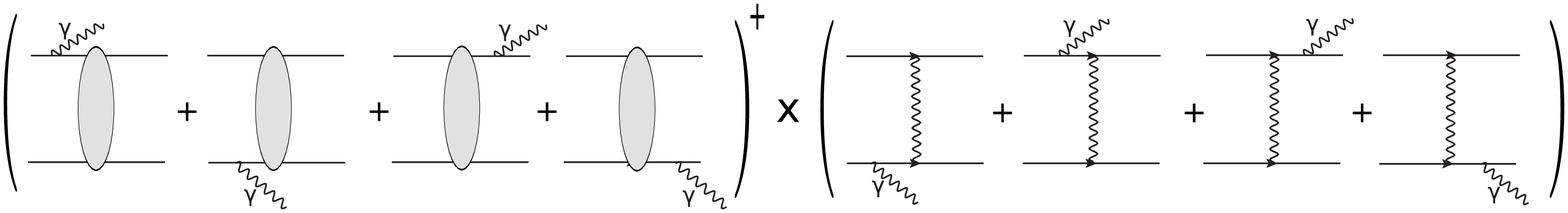}
\par\end{centering}
\begin{centering}
\includegraphics[scale=0.5]{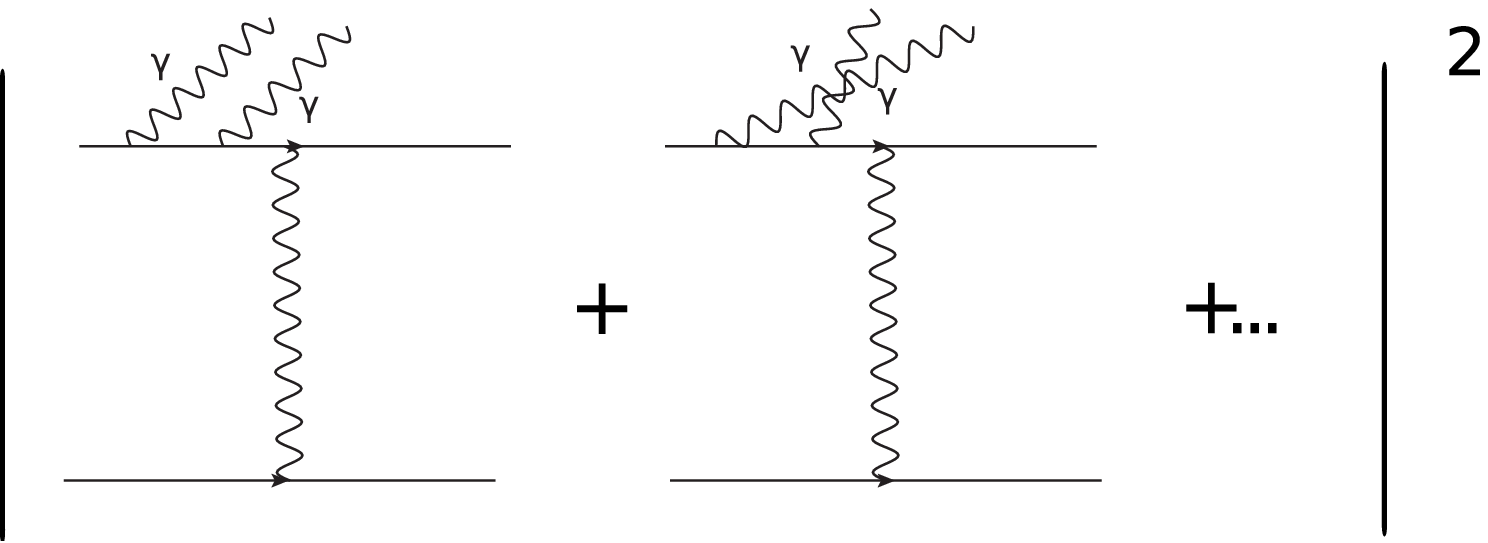}
\par\end{centering}

\caption{Bremsstrahlung treatment of IR divergences in the Q-part. The top
plot (a) represents interference between emission from one-loop (shaded
bubble) and Born graphs. The bottom plot (b) is the two-photon emission amplitude squared.}

\label{two-loops-bremss}
\end{figure}

 The higher-order corrections ($\propto\alpha^{4}$) to the electroweak
Born cross section can be divided into two classes, Q-part and T-part. The Q-part is
induced by the quadratic one-loop amplitude ($\sim M_{1}M_{1}^{\dagger}$)
(third term in Eq.~\ref{eq:a3}) and the T-part is an interference term between the Born and two-loop amplitudes: $\sigma_{T}=\frac{\pi^{3}}{s}\mbox{Re}M_{2}M_{0}^{\dagger}\propto\alpha^{4}$
(Fig.~\ref{two-loops-Moller}).
The T-part still needs to be evaluated in the future, but we can provide
some results for the Q-part in this paper. The cross section
for the Q-part can be divided into two terms: 
$\sigma_{Q}=\sigma_{Q}^{\lambda}+\sigma_{Q}^{f}$.
The first term, $\sigma_{Q}^{\lambda}$, is an IR-divergent and regularized
part of the cross section and the second term, $\sigma_{Q}^{f}=(\frac{\alpha}{\pi})^{2}\delta_{1}^{f}\cdot\sigma_{0}$, is a finite
contribution. The IR-divergent part has the following
structure:
\begin{eqnarray}
\sigma_{Q}^{\lambda}=\frac{\pi^{3}}{2s}M_{1}^{\lambda\dagger}(M_{1}^{\lambda}+2M_{1}^{f})=\frac{1}{4}\Big(\frac{\alpha}{\pi}\Big)^{2}\mbox{Re}\Big[\delta_{1}^{\lambda*}(\delta_{1}^{\lambda}+2\delta_{1}^{f})\Big]\cdot\sigma_{0},\label{eq:a10}
\end{eqnarray}
where $\delta_{1}^{\lambda}=4\ln\frac{\lambda}{\sqrt{s}}\Big(\ln\frac{tu}{m^{2}s}-1+i\pi\Big)$. 
Since the Q-part contains terms of order $\propto\ln^{2}\frac{\lambda}{\sqrt{s}}$
it deserves a special attention. To treat the IR divergences,
we have to account not only for photon emission from one-loop diagrams
but also include a complete treatment of the two-photon emission (Fig.~\ref{two-loops-bremss}).
A half of the bremsstrahlung contribution in Fig.~\ref{two-loops-bremss}(a)
and (b) is responsible for the treatment of IR divergence in the Q-part and
the other half for the T-part. We take the maximum energy of the emitted soft photon to be $\omega=0.05\cdot\sqrt{s}$.
The bremsstrahlung cross section for Q-part is derived from the Fig.~\ref{two-loops-bremss}(a)
as: 
\begin{eqnarray}
\sigma_{Q}^{\gamma}=\frac{1}{2}\sigma^{\gamma}=\frac{1}{2}\Big(\frac{\alpha}{\pi}\Big)^{2}\mbox{Re}[(-\delta_{1}^{\lambda}+R_{1})^{*}(\delta_{1}^{\lambda}+\delta_{1}^{f})]\cdot\sigma_{0}\label{eq:a11}\\
R_{1}=-4\ln\frac{\sqrt{s}}{2\omega}\Big(\ln\frac{tu}{m^{2}s}-1+i\pi\Big)-\ln^{2}\frac{s}{em^{2}}+1-\frac{\pi^{3}}{3}+\ln^{2}\frac{u}{t}.\nonumber 
\end{eqnarray}
Here, $\sigma^{\gamma}$ is the total photon emission cross section and
$\sigma_{Q}^{\gamma}$ is the one-photon bremsstrahlung term from the Q-part.
The two-photons emission for the Q-part ($\sigma_{Q}^{\gamma\gamma}$)
is derived from Fig.~\ref{two-loops-bremss}(b):
\begin{eqnarray}
\sigma_{Q}^{\gamma\gamma}=\frac{1}{2}\sigma^{\gamma\gamma}=\frac{1}{4}\Big(\frac{\alpha}{\pi}\Big)^{2}\Bigg(\Big|-\delta_{1}^{\lambda}+R_{1}\Big|^{2}-\frac{8}{3}\pi^{2}\Big|\ln\frac{tu}{m^{2}s}-1+i\pi\Big|^{2}\Bigg)\cdot\sigma_{0}.\label{eq:a12}
\end{eqnarray}
Combining Eqs.~\ref{eq:a10}, \ref{eq:a11} and \ref{eq:a12}
gives the final result for $\sigma_{Q}^{\lambda}+\sigma_{Q}^{\gamma}+\sigma_{Q}^{\gamma\gamma}$
free from nonphysical parameters with the regularization parameter $\lambda$ cancelled analytically. Detailed calculations can be found in \cite{q-part}. 
\begin{figure}
\begin{centering}
\includegraphics[scale=0.48]{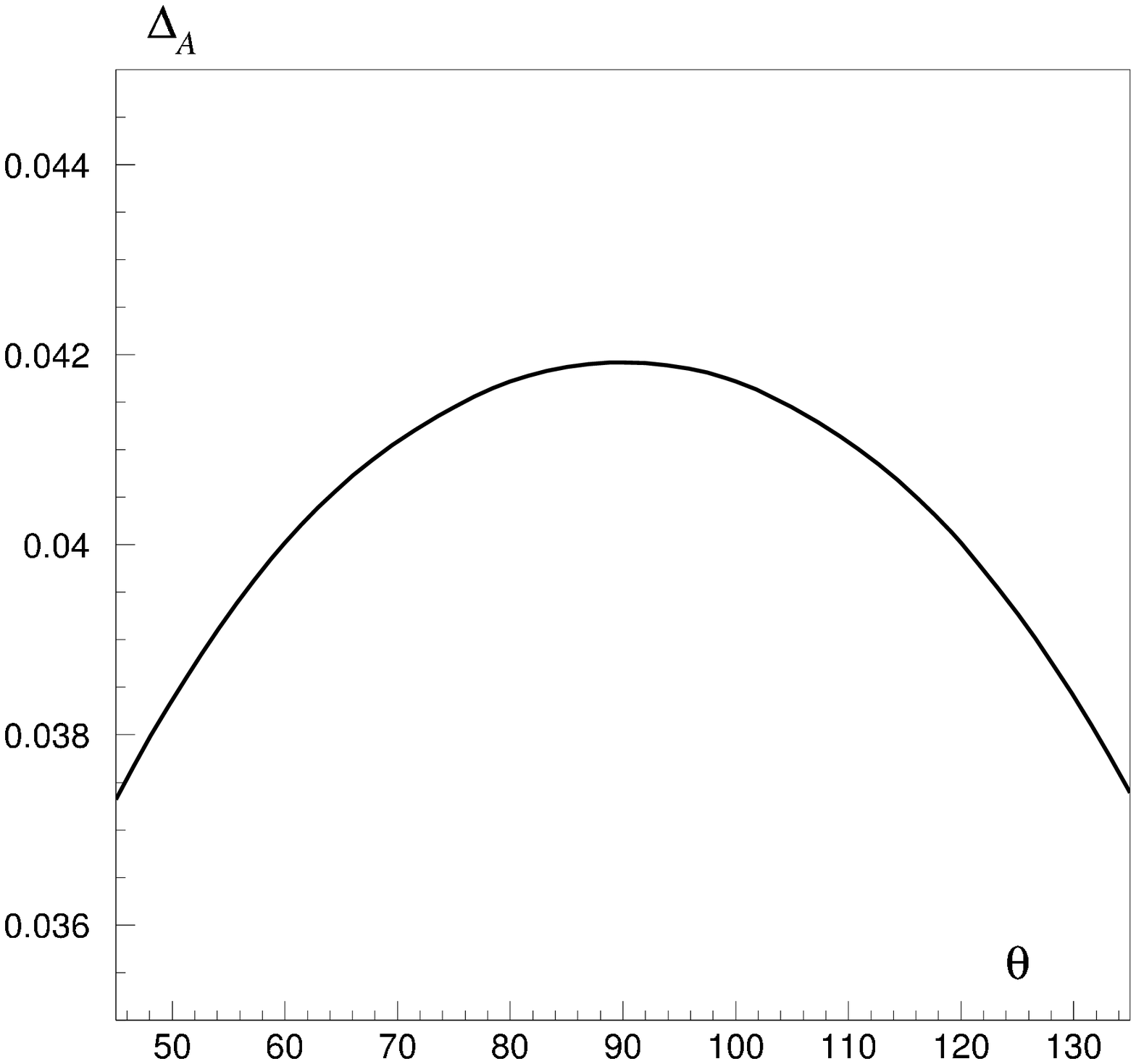}~~~
\includegraphics[scale=0.45]{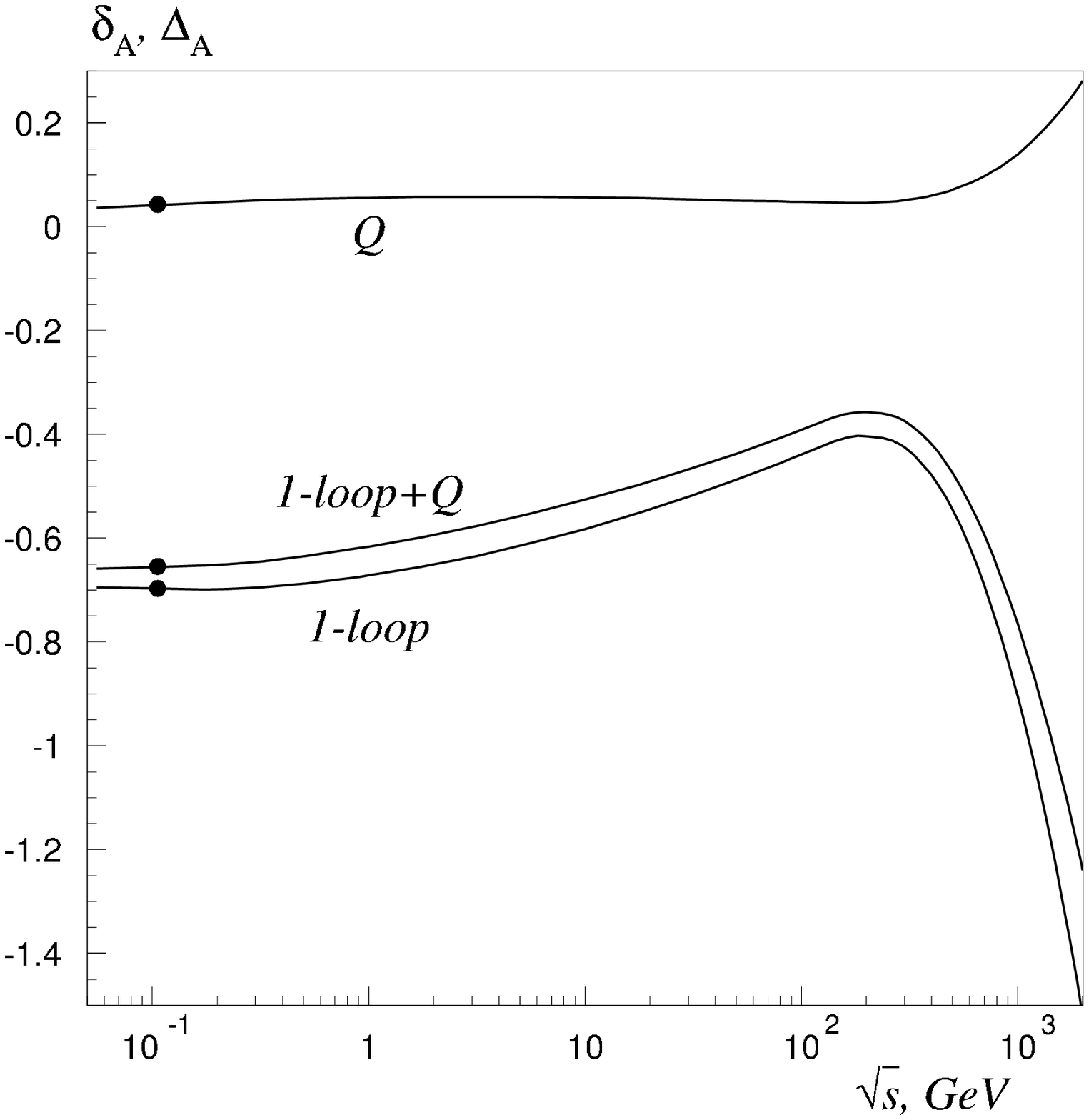}
\par\end{centering}

\caption{The left plot (a) shows the angular dependence
of the correction induced by the quadratic part only ($\Delta_{A}$) and the right plot
(b) shows the energy dependence of the ``one-loop'', ``one-loop + Q-part'' and quadratic corrections.}

\label{q-part}
\end{figure} 

As one can see from Fig.~\ref{q-part}(a),
the correction induced by the Q-part
($\Delta_{A}=(A_{LR}^{1-loop+Q}-A_{LR}^{1-loop})/A_{LR}^{0}$) can reach as much as 
$\sim4\%$ at $90^{\circ}$. The energy dependence (Fig.~\ref{q-part}(b)) is nearly constant for
$\sqrt{s}<m_{Z}$ but increases rapidly after weak interactions
become comparable to QED. 

\section{Conclusion}
With the one-loop corrections now under control, it is worth considering the electroweak radiative corrections at the two-loop level. 
One way to find some indication of the size of higher-order contributions is to compare results that are expressed 
in terms of quantities related to different renormalization schemes, and our tuned comparison between the results 
obtained in the on-shell and constrained differential renormalization schemes show a difference of about 11\%. Although an argument can be made that the two-loop 
corrections are suppressed by a factor of $\alpha\pi$ relative to the one-loop corrections, we believe that they can
 no longer be dismissed, especially in the light of the 2\% uncertainty to asymmetry promised by the MOLLER experiment. 
At the MOLLER kinematic conditions, the part of the quadratic correction considered here can increase the asymmetry 
up to $\sim4\%$. For the high-energy region $\sqrt{s}\sim2\,\mbox{TeV}$, a contribution from the quadratic correction 
can reach +30\%. It is impossible to say at this time if the Q-part will be enhanced or cancelled by other two-loop 
radiative corrections, but we believe that its size demands a detailed and consistent consideration 
of the T-part, which is the current task of our group.

\section{ACKNOWLEDGMENTS}

We are grateful to Y. Bystritskiy and T. Hahn
for stimulating discussions.
A. A. and S. B. thank the Theory Center at Jefferson Lab, and V. Z. thanks
Acadia University for hospitality in 2011. This work was supported by the
Natural Sciences and Engineering Research Council of Canada
and Belarus scientific program "Convergence".

\end{document}